# Decomposition and utilization of source and receiver ghosts in marine seismic reflection survey data

Hiroaki Ozasa,[1] and Hitoshi Mikada[2]

[1] Intelligent Information Management Headquarters, IHI Corporation, Koto-ku, Tokyo 135-8710, Japan; ozasa3622@ihi-g.com

[2] Department of Civil and Earth Resources Engineering, Kyoto University, Kyoto-shi, Kyoto 615-8540, Japan; mikada@gakushikai.jp

In marine seismic reflection surveys, most data comprise only the pressure acquired by a hydrophone array. The acquired data are subject to frequency bandwidth limitations caused by the contamination of surface-reflected ghost signals associated with seismic signals generated by artificial sources such as airguns. This study developed a method to exploit the mirror-image effect to control these signals and image the subsurface structure using signals generated by virtual seismic sources at locations mirroring actual sources. The processing results indicate that decomposed ghost signals can be regarded as additional survey data to enhance the signal-to-noise ratio and widen bandwidth.

1. **Introduction**

The use of broadband seismic data in offshore exploration has become possible primarily because of advances in hardware over the past decade. A significant obstacle to acquiring this broadband data was the superposition of sea-surface reflections, also known as "ghost signals," generated by the upcoming source signal from seismic sources and by downgoing surface-reflected waves to receivers, respectively. Previous studies have investigated methods to remove the receiver ghost using sophisticated hardware (e.g., Amundsen and Reitan, 2014; Moldoveanu et al., 2007; Robertsson et al., 2008; Tenghamn et al., 2007) to mitigate the effects of the receiver ghost. In addition, synchronized multiple depth sources have been used to eliminate the source ghost, like the technique used for over-under streamer cables (Siliqi et al., 2013). The main objectives to deploy sophisticated data acquisition system include to enhance the coverage at frequency notches that appear as a function of the towing depths of sources and receivers. However, data acquisition using the latest hardware is not always possible, and there is a massive amount of legacy exploration data assets acquired by conventional marine seismic systems. The deghosting of marine seismic data acquired only by pressure sensors has been actively undertaken (Denisov et al., 2018). Sophisticated methods have been applied using inverse modeling (Cao and Blacquière, 2021), wave-equation-based data processing (Amundsen et al., 2013), a mixture of physical modeling, and inversion (Denisov et al., 2018), and machine learning technologies (Vrolijk and Blacquière, 2020). These methods succeeded in suppressing ghost signals superimposed on the acquired data. However, none of them have attempted to utilize the decomposed ghost-signal energy. It has been shown that sea-surface reflections separated from the acquired wavefields can be considered as data from another set of seismic source/receiver arrays located in a mirror-image position to the sea surface to exploit the efficient reflections at the water-air interface (e.g., Lu et al., 2015; Jamali H. et al., 2019). Lu et al. (2015) and Jamali H. et al. (2019) utilized the acquired data with dual-sensor streamers and a vertical cable seismic method, respectively, to decompose upcoming and downgoing waves from acquired data. The source ghost signal can be regarded as a signal from a virtual source at the mirror-image point of the actual towed source with respect to the sea surface. Similarly, it can be assumed that the receiver ghost acts as a wave incident on hypothetical receivers at the mirror-image points of the actual receivers. We hypothesized that the mirror-image concept could realize the configuration of sources and streamers at multiple depths that compensate for frequency notches caused by the ghosts in this study.

This study applied the mirror-image concept to utilize source and receiver ghost signals. First, an algorithm was developed to separate upward-propagating waves incident on the streamer from downward-propagating waves from the sea surface for each common shot gather; the algorithm was developed under the assumption that waves incident to the streamer from the sea surface were virtually incident to the mirror-imaged streamer towed above the sea surface. Since the downward traveling waves incident on the streamer could be considered as waves incident to the mirror-imaged streamer above the sea surface. The configuration therefore would become the same as over/under data acquisition (Moldoveanu et al., 2007). After the decomposition of waves virtually incident to the mirror-imaged streamer, the waves would be downward-continued to the actual streamer location and stack with the upcoming-only waves incident to the streamer to improve the signal-to-noise ratio (SNR) by the square root of two. The application of reciprocity to exchange a source and a receiver would require the same algorithm for each of common receiver gathers. The sources towed beneath the surface, and their mirror-imaged virtual sources would form virtual over/under streamers or multilevel sources (Fromyr et al., 2009). Data processing same as for common shot gathers improves the SNR by the square root of two for the second time. Therefore, the procedure to use the mirror image on both source and receiver sides improves the SNR by two. We conducted a sea trial with a 7.87-L (480 in$^3$) airgun array, and a 2,300-m-long conventional streamer cable towed at a depth of 6 m to verify our hypothesis. This study attempted to separate the ghost signals by data processing and evaluate the difference in the frequency band and the SNR of the data before and after processing. Once our hypothesis is justified, the frequency range of marine seismic reflection would be extended, which may lead to the use of full waveform inversion (FWI) of legacy data and improve accuracy in time-lapse reservoir characterization (Kneller et al., 2013).

**2. Data Processing Method**

Predictive deconvolution has been proposed to shorten the time length of the seismic wavelet in the acquired seismic records (Peacock and Treitel, 1969). The method uses the waveform of the primary signals and the periodicity of the reverberations to attenuate the reverberation. However, predictive deconvolution is not designed to remove sea-surface reflection, which comprises waves that travel upward from the seismic source and are reflected only once at the sea-surface. Moreover, as Roberts and Goulty (1988) reported, it is difficult to consider the directional dependence of the output waveform caused by the difference in distance from the seismic source to the sea-surface reflection point. Thus, it is necessary to employ a new method different from the conventional multiple-removal method. It is known that upward traveling waves from an underwater seismic source reach the sea surface at different times depending on their incident angle and that periodicity-dependent processing results in insufficient separation of wavefields (Taner, 1980). Although the impaired periodicity can be compensated for by multichannel predictive deconvolution (Taner, 1980), further accuracy is necessary for high-precision data processing. In this study, sea-surface reflections were identified using a method that reproduces the physical process of generating reflected waves.

*2.1 Removal of sea-surface reflection*

This section describes the processing method used to remove the receiver ghost. A similar concept is applied to remove the source ghost using the reciprocity concept, as shown schematically in Fig. 1. First, we begin with the following equation:

$$k_x^2 + k_y^2 + k_z^2 = \frac{\omega^2}{v^2}, \qquad (1)$$

where $k_x$, $k_y$, and $k_z$ denote the propagation vectors in the x-, y-, and z-axes, respectively, as shown in Fig. 1. $\omega$ denotes the angular frequency, and $v$ the water sound velocity. We derived the following relation from Eq. (1) as follows:

$$k_z = \pm \frac{\omega}{v}\sqrt{(1 - \frac{k_x^2+k_y^2}{\omega^2})}. \qquad (2)$$

Therefore, the upcoming wavefields P(x,y,z=d) incident on the virtual receivers are expressed by the one-way wave equation.

$$\frac{\partial}{\partial z}P = i\frac{\omega}{v}P\sqrt{(1 - \frac{k_x^2+k_y^2}{\omega^2})}. \qquad (3)$$

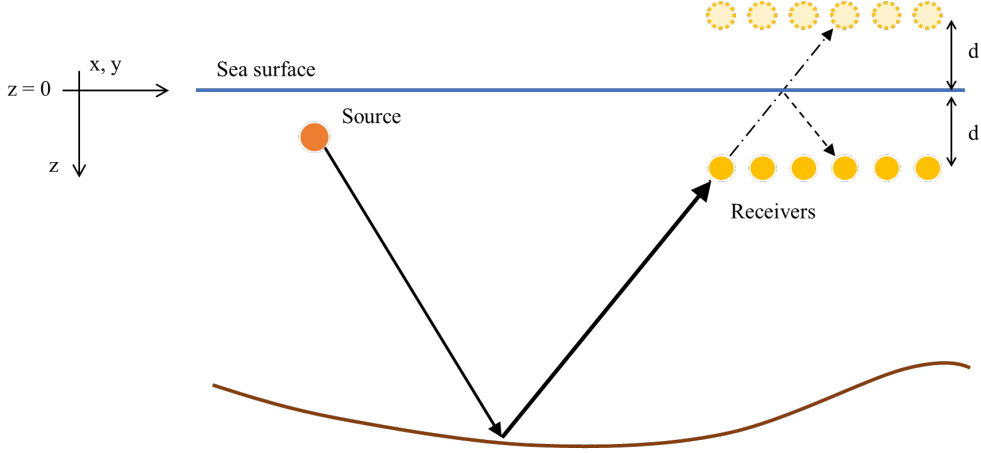

Fig. 1. Schematic of wavefields acquired by receivers. The downgoing wavefield, reflected at the sea surface, can be regarded as the upcoming wavefield acquired by the virtual receivers located in a mirror-imageposition to the sea surface $(z = -d)$.

The wavefields acquired at $z = -d$ are expressed assuming that the wavefields $P(x, y, z = -d, \omega)$ comprise only upcoming waves, and that the mirror reflection occurs in the upcoming waves at the sea surface.

$$P(x, y, z = -d, \omega) = -i \int_d^{-d} \left\{ \frac{\omega}{v} P(x, y, z = d, \omega) \sqrt{\left(1 - \frac{k_x^2 + k_y^2}{\omega^2}\right)} \right\} dz.$$

(4)

The upcoming wavefields that $P_0$ would virtually arrive at $z = -d$ for an unbounded medium, are reflected at the sea surface $(z = 0)$ to become downgoing wavefields traveling to $z = d$. Therefore, the wavefields at $z = d$ without sea-surface reflection $\widetilde{P}_0(x, y, z = d, \omega)$ can be expressed as follows:

$$\widetilde{P}_0(x, y, z = d, \omega) = P(x, y, z = d, \omega) - r^{(1)}(k_z, \omega) P_0(x, y, z = -d, \omega),$$

(5)

where $r^{(1)}(k_z, \omega)$ represents the incident-angle-dependent reflection coefficient (IRC) at the sea surface for the first iteration. The wavefields $\boldsymbol{P}$ are expressed by the superposition of the upcoming wavefields $\boldsymbol{P_0}$ and the sea surface reflected $\boldsymbol{P_0}$ as follows:

$$\boldsymbol{P}(x, y, z = d, \omega) = \boldsymbol{P_0}(x, y, z = d, \omega) + r^{(1)}(k_z, \omega)\boldsymbol{P_0}(x, y, z = -d, \omega).$$

(6)

Equation (6) may be transformed into a recursive form as:

$$\boldsymbol{P_0}(x, y, z = d, \omega) = \boldsymbol{P}(x, y, z = d, \omega) - r^{(1)}(k_z, \omega)\boldsymbol{P_0}(x, y, z = -d, \omega) \quad (7)$$

Equation (7) is a formula that recursively estimates the upcoming wavefields $\boldsymbol{P_0}$ from the observed wavefields. The following equation is used to estimate the second term of Eq. (7).

$$\boldsymbol{P_0}(x, y, z = -d, \omega) = \boldsymbol{P}(x, y, z = -d, \omega) - r^{(2)}(k_z, \omega)\boldsymbol{P_0}(x, y, z = -3d, \omega),$$

(8)

where $\boldsymbol{P}(x, y, z = -d, \omega)$ and $\boldsymbol{P_0}(x, y, z = -3d, \omega)$ are the observed wavefields extrapolated at $z = -d$ assuming $\boldsymbol{P}$ is composed only of the upcoming wavefields, and the downgoing waves reflected at the virtual sea surface located at $z = -2d$. The following equation holds after the iteration for $n$ iterations.

$$\boldsymbol{P_0}(x, y, z = -(2n - 1)d, \omega)$$
$$= \boldsymbol{P}(x, y, z = -(2n - 1)d, \omega) - r^{(n)}(k_z, \omega)\boldsymbol{P_0}(x, y, z = -(2n + 1)d, \omega)$$

(9)

Equations (8) and (9) can be used to obtain the following equation.

$$P_0(x, y, z = d, \omega)$$

$$= \sum_{m=1}^{n-1} \prod_{l=1}^{m} (-1)^l r^{(l)}(k_z, \omega) P(x, y, z = -(2m-1)d, \omega)$$

$$+ \prod_{l=1}^{n} (-1)^l r^{(l)}(k_z, \omega) P_0(x, y, z = -(2n-1)d, \omega)$$

(10)

When $(2N-1)d/v$, where $v$, the water sound velocity, exceeds the maximum recording time after the $N$-th iterations, the second term of Eq. (10) becomes negligible. The following equation expresses the final wavefields without ghost-signal contamination in the recording time range:

$$\widetilde{P}_0(x, y, z = d, \omega)$$

$$= P(x, y, z = d, \omega) + \sum_{m=1}^{N-1} \prod_{l=1}^{m} (-1)^l r^{(l)}(\mathbf{k_z}, \boldsymbol{\omega}) P(x, y, z = -(2m-1)d, \omega)$$

(11)

Equation (11) estimates the wavefields at $z = d$ without sea-surface reflection $\widetilde{P}_0$ using only the observed wavefields $P$ that contain both the upcoming and downgoing ward wavefields. Our processing method eliminates the influence of the ghosts in a finite number of iterations only by extrapolating the initially acquired data $P(x, y, z = d)$ every $2d/v$ time-step and estimate the IRC $r^{(l)}(\mathbf{k_z}, \boldsymbol{\omega})$ in Eq. (11) for $N$ times.

The signal strength and depth of the receivers may vary slightly owing to sea-surface roughness, and the reflection coefficient $r(k_z, \omega)$ was introduced in Eq. (5). Cecconello et al. (2018) and Konuk and Shragge (2020) discussed that the reflection coefficient of the sea surface for sound waves may not always be of a flat mirror because the flatness varies as a function of time and space. In this study, the IRC are estimated to compensate for the sea-surface roughness effects with high efficiency from the acquired data $P(x, y, z = d)$ and obtain the wavefield at $z = d$ without sea-surface reflection $P_0$ instead of Eq. (7).

$$P_0 = P - H'f', \qquad (12)$$

where $H'$ is a matrix of the transposed virtual upcoming wavefield $P_0(x, y, z = -(2m-1)d)$, and $f'$ denotes a matrix of multichannel filters equivalent to the IRC. The iterative shrinkage-thresholding algorithms (FISTA; Beck and Teboulle, 2009; Li et al., 2016; Liu and Lu, 2015) was applied in this study to calculate the prediction filter. Using FISTA and $\ell_1$ regularization approach (Beck and Teboulle, 2009), the prediction filter $f'$, satisfying the following equation, was obtained:

$$\text{Arg } \min_{f'}\{\|P - H'f'\|_2^2 + \lambda\|f'\|_1^2\}, \qquad (13)$$

where $\|\cdot\|_n$ denotes the $\ell_n$ norm and $\lambda$ is a hyperparameter indicating the influence of the regularization term. In this study, $\lambda = 0.001$ was chosen as per Li et al. (2016). Based on the criterion in Eq. (13), seven traces with seven time-samples/trace were chosen to represent the IRC. Since the IRC could be a function of time during the data acquisition, IRC are estimated at every iteration as indicated in Eq. (11). After the iterations for N-times, the second term of Eq. (11) was obtained as:

$$P_0(x, y, z = d, \omega) = \sum_{m=1}^{N-1} \prod_{l=1}^{m}(-1)^l r^{(1)}(k_z, \omega) P(x, y, z = -(2m-1)d, \omega). \qquad (14)$$

Therefore, we can obtain the direct wave component $\widetilde{P}_0$ using Eq. (11). Since both upcoming and downgoing wavefields from the sources and those to the receivers were estimated, the configuration of virtual multilevel sources and virtual over and real under streamers is actualized to compensate the influence of frequency notches caused by ghost signals.

*2.2 Mirror image processing (Utilization of reflection)*

This section discusses a method for utilizing a separated sea-surface reflection wave as a signal component. The sea-surface reflection can be regarded as survey data by a hypothetical survey system at the mirrored location of the actual system with respect to the sea-surface. Therefore, we attempted to improve the signal strength of the survey data by adding the separated sea-surface reflection data to the direct wave data in the capacity of survey data acquired by a mirrored survey system. Fig. 2 shows the conceptual diagram for the method. Under this method, a single seismic source can acquire the same amount of survey data as two sources— an actual source and a mirror-image source.

The mirror-image data had a delay of $t = 2d/v$ from the direct survey data (Fig. 2). Therefore, we employed upward propagation processing of $t = 2d/v$ for the mirror-image data $P(x, y, z = -d)$ to correct the difference between the direct survey data and mirror-image data caused by the difference in propagation paths. It is expected that the corrected mirror-image data can be incorporated to direct survey data with high accuracy. The survey data of four different propagation paths can be acquired by decomposing both the source and receiver ghosts from the acquired data. Combining these four datasets would quadruple the signal strength of the survey data and thus double ($=\sqrt{4}$) the SNR.

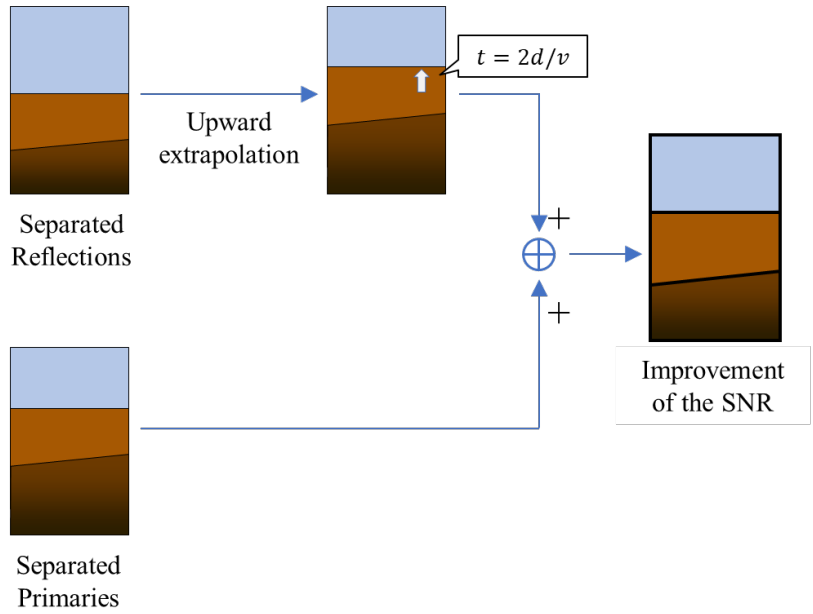

Fig. 2. Conceptual diagram of signal enhancement using separated sea-surface reflection. Upward extrapolation processing with time delay $t = 2d/v$ is applied to the separated sea-surface reflection data, i.e., the downward continuation of either the source or receiver, to eliminate the effect of propagation path differences. After location alignment of both the source and mirrored source or the receiver and mirrored receiver, the primaries and the separated sea-surface reflections are accurately stacked.

## 3. Results

We employed a data processing method to the survey data acquired by the survey system consisting of a 7.87-L (480 in³) airgun array towed at a depth of 6 m and a 2,300-m-long streamer cable with 168 channels towed at 6 m. We processed the acquired data to verify our hypothesis pertaining to the decomposition and utilization of ghosts. During processing, the 15° wave-extrapolation formula (Claerbout, 1985) was applied to the wavefield extrapolation. Fig. 3 shows the survey images before and after the data processing. Fig. 4 (a) shows the improvement achieved in signal strength and SNR by utilizing separated sea-surface reflection. As per our estimations in Section 2.1, the SNR was improved by approximately two times (Table 1). The frequency spectrum before the separation of the sea-surface reflection (Fig. 3 (a)) and after the utilization of sea-surface reflection (Fig. 3.b) are shown in Fig. 4 (b). It can be observed that our processing restores the amplitudes in the frequency notches at low and high frequencies.

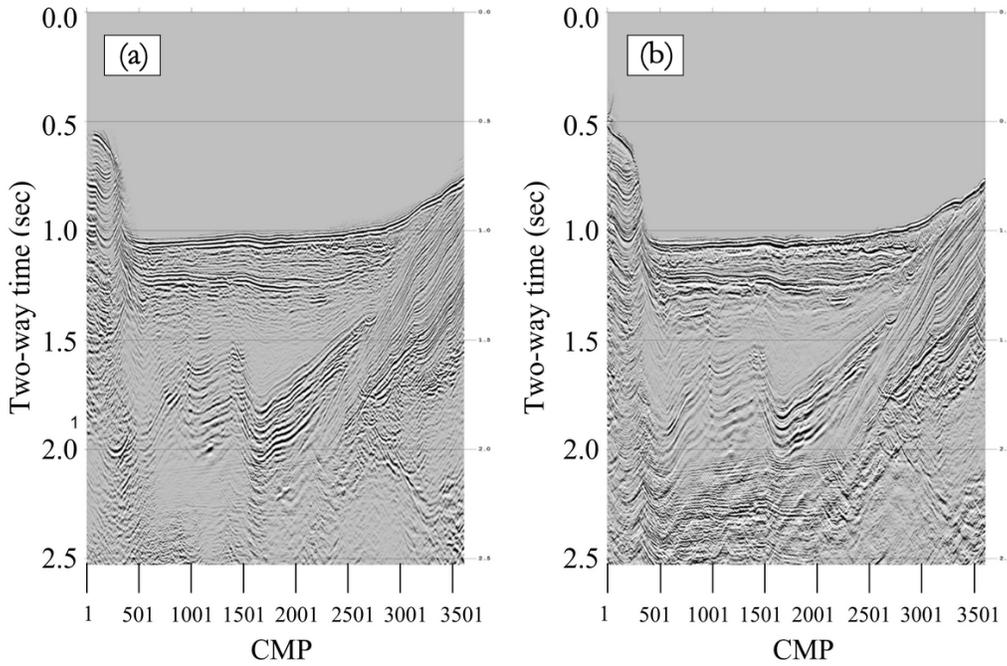

Fig. 3. Migrated images of the sea trial data acquired by an airgun array and a towed streamer (a) after the conventional processing, and (b) after the proposed processing using mirror-image. Deepwater reverberations around the time depths from 2.0 to 2.5 s buried in the noise in the left image have become amplified in the right image due to the signal-to-noise ratio enhancement by a factor of two. The common

middle point (CMP) interval is 6.25 m. We employed the reverse time migration with the same velocity model to both data sets.

TABLE 1. SNR of CMP 1701 of Fig. 3. Noise level is represented as root mean square (RMS) level between t = 0.9 to 1.0 s. Signal level is also represented as RMS between t = 1.0 to 1.2 s, where t denotes the two-way time. The combination of four data sets with different paths, i.e., direct and reflected waves at both source and receiver sides, has improved the signal level of the processed data to approximately four times the signal level of the original data. Consequently, the SNR has been approximately improved by a factor of two (=78.5 / 41.4). Note that all the trace amplitudes were normalized to a constant value and the numbers for noise and signal levels are in the arbitrary unit (a.u.).

|  | Noise level | Signal level | SNR |
|---|---|---|---|
| Original data | 0.00193 | 0.0797 | 41.4 |
| Processed data | 0.00334 | 0.262 | 78.5 |

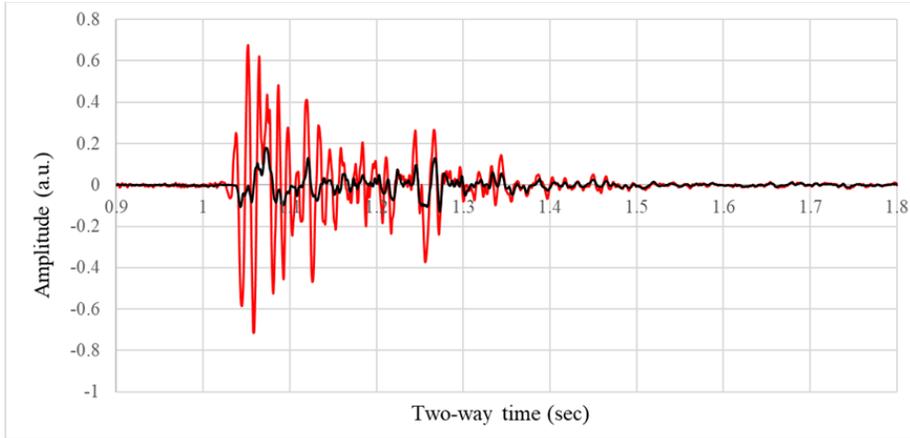

(a)

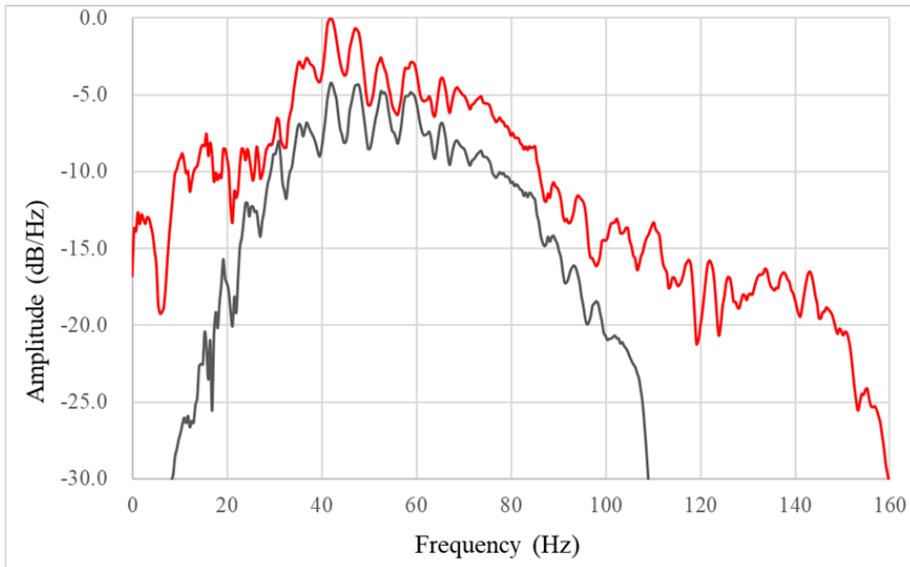

(b)

Fig. 4. Survey results at CMP 1701 in the migrated images in Fig. 3. The black and red solid lines represent the waves without and with mirror-image processing, respectively. (a) Result in the time domain. The summation of the sea-surface ghosts at both source and receiver sides exaggerates the amplitudes in the mirror-image processing. (b) Result in the frequency domain. There is a frequency notch at 125 Hz caused by the towing depth of the airgun array and the streamer cable, 6 m. The frequency bandwidth damaged by the sea-surface reflection in the original data has been recovered in the processed data.

**4. Discussion**

Predictive deconvolution filters (Peacock and Tritel, 1969) have removed periodically introduced noise. For example, the bubble effect of an air gun or the removal of multiple reflections between the water surface and the seabed in an ocean with smooth depth variation. It has recently been used as a multi-channel filter predictive deconvolution to deal with sea surface multiple reflections in subsurface structures with lateral variations (Li et al., 2016; for example). However, seismic source ghosting and receiver ghosting have not been addressed. As for the prediction lag, a feature of predictive deconvolution filters, a simple time shift has been taken account of, but the effect of wave propagation has not been well taken into consideration, except for multi-channel data. The method used in this study differs from the conventional multi-channel prediction type convolution only. The prediction lag is considered the wave propagation time, and wave extrapolation is introduced instead of time shift. Our results show that it is necessary to take into account the effect of wave propagation when extending the originally one-dimensional signal processing of deconvolution filters to multiple dimensions. More importantly, our results demonstrated that the idea of prediction distance to be extended to extrapolation time to deal with both source and receiver ghost signals.

Since Haggerty (1956) and Pavey and Pearson (1966) patented a method for removing the effect of receiver ghost signals by deploying hydrophones and geophones or over/under streamer cables, various hardware technologies based on similar concepts, such as the use of hydrophone-geophone streamers (e.g., Tenghamn et al., 2007), multicomponent streamers (e.g., Robertsson et al., 2008), over/under streamers (e.g., Moldoveanu, et al., 2007), and curved streamers (e.g., Amundsen and Reitan, 2014), have been realized. These technologies have succeeded in eliminating the effects of receiver ghosts from the acquired survey data to widen the frequency band of the data. However, all the above mentioned methods require sophisticated systems, and no solution to survey data acquired by conventional systems, including legacy data, has been presented.

Ghosts caused by sea-surface reflections interfere with direct waves, producing notches in the frequency spectrum depending on the deployed depth of the sources or receivers. The produced frequency notch suppressed the energy in a wide range of frequencies. Moreover, low frequencies are especially crucial for FWI because low frequencies can deliver accurate velocity models and provide reliable reservoir inversion results (e.g., Jupp et al., 2012; Mothi et al., 2013). As the approaches mentioned earlier have been applied to remove receiver ghosts, applying a synchronized multilevel source array has also been attempted on the source side (Cambois et al., 2009). However, multilevel source technology is not applicable in survey data acquired by conventional systems, and the problem of source ghosts requires processing solutions for single-depth seismic sources. It is also advantageous to design processing methods that broaden the frequency contents of signals traveling downward from sources without source ghosts using conventional survey instruments and methods. Moreover, the resolution of seismic time-lapse surveys that use legacy data as reference data would increase if the reference data became broadband.

The problems associated with ghost signals overlapped with the primary source signals have been left untouched under the assumption that ghosts are inseparable from the source signal; the only exception has been the development of hardware. This study shows that the surface-reflected ghosts can be decomposed by accurately estimating the distance between sources or receivers and the sea surface that causes the travel time difference between the downgoing and upcoming waves in the vicinity of sources or receivers, assuming that the sea surface is acoustically close to a mirror surface. When the downgoing and upcoming waves from sources or to receivers are separated by the proposed method, the source and receiver ghosts are considered signals from mirror-imaged sources or to mirror-imaged receivers located at different depths. In other words, it would be possible to realize a virtual multilevel source array and virtual over/under streamer cables. As we have already verified, this method can be applied to remove frequency notches and widen the bandwidth even for pressure-only acquisition data, that is, marine seismic reflection data acquired with conventional survey systems, including already-acquired legacy data. Furthermore, it can enhance the signal-to-noise ratio of the acquired data by a factor of two using the ghost signals after the downward continuation of decomposed data for the mirror-imaged source and receivers. An exclusive prerequisite is the availability of positional data for shot points and streamer locations in the three-dimensional space generally recorded during data acquisition.

## 5. Conclusion

We verified the effectiveness of ghost de-blending processing based on wavefield extrapolation processing, which was designed for long-period sea-surface reflection. The study results indicate the following:

(1) Frequency notches, which appear as a function of the towing depth of the seismic source or the receiver, can be recovered via processing to separate the source/receiver ghost. Thus, it was possible to achieve a broadband frequency range for the acquired data. The extension to lower frequencies could be advantageous for applying FWI to marine seismic data acquired by conventional data acquisition systems.

(2) The separated ghost can be treated as a seismic signal from a mirrored source or to a mirrored receiver to improve the signal-to-noise ratio of the seismic data.

(3) The proposed mirror-image processing method compensates for the effects of the towing depths of the source and receiver in marine reflection surveys. Thus, towing instruments deeper than conventional operational depths may become possible.

Specific instruments are required to remove the ghost signals and resulting broadband. Likewise, the proposed method enhances the bandwidth for data acquired by conventional instruments, including legacy marine seismic reflection data. Enhanced bandwidth would surely make use of legacy data for time-lapse reservoir characterization using FWI.

**Acknowledgement**

We would like to thank Drs. E. Jamari Hondori, Junichi Takekawa, and Eiichi Asakawa, and Ms. Shiori Kamei for valuable discussion and Editage (www.editage.com) for English language editing.